\documentclass[twocolumn,preprintnumbers,amsmath,amssymb]{revtex4}

\usepackage{graphicx}
\usepackage{dcolumn}
\usepackage{bm}

\begin{document}


\title{Positronium lifetime in polymers}
\author{Abel Camacho}
\email{acq@xanum.uam.mx}
\affiliation{Department of Physics \\
Universidad Aut\'onoma Metropolitana--Iztapalapa\\
Apartado Postal 55--534, C.P. 09340, M\'exico, D.F., M\'exico.}
\date{\today}

\begin{abstract}
A model describing the relationship between the ortho--positronium
lifetime and the volume of a void, located in a synthetic zeolite,
is analyzed. Our idea, which allows us to take into account the
effects of temperature, comprises the introduction of a
non--hermitian term in the Hamiltonian, which accounts for the
annihilation of the ortho--positronium. The predictions of the
present model are also confronted against an already known
experimental result.

Key words: Positronium lifetime, polymers
\end{abstract}
\maketitle

\section{Introduction}
\bigskip

In the context of polymer research one of the most interesting
issues is connected with the relationship between molecular
relaxation and free--volume properties \cite{[1]}. Experimentally
several techniques have been already developed in order to
determine the size of these voids in polymers, for instance,
small--angle X ray diffraction \cite{[2]}, neutron diffraction
\cite{[3]}, or positron annihilation spectroscopy \cite{[4]}, in
which the positron or positronium play the role of a probe, and
the lifetimes of these systems are monitored.

The current experimental results give evidence that the positron
and positronium are localized in the pre--existing free--volumes
in polymers \cite{[5]}, and in consequence one could relate the
size of the free--volume region with the detected lifetime
\cite{[6]}.

The use of positron or positronium lifetimes to determine the size
of voids faces severe difficulties \cite{[1]}, which leave only
the possibility of employing approximated a\-pproaches to
determine the correlation between lifetime and free--volume.
Clearly this is an indirect method, and in consequence we need a
model to link the measured parameter with the required variable.
Usually this requirement has been satisfied employing the so
called standard model \cite{[7]}. Nevertheless it must be
underlined that, as already noted \cite{[7]}, this model embraces
several deficiencies.

In the present work it will be assumed that the positronium is
located in a sphe\-rical wall, but, in order to avoid one of the
existing drawbacks (to wit, that the matrix electrons form a layer
of homogeneous density and thickness), we will not only consider a
finite wall, but also we will introduce a non--hermitian term in
the Hamiltonian, which accounts then for the annihilation of the
positro\-nium due to its interaction with the polymer. In this way
the atomic structure of the polymer is taken into account in a
more realistic ma\-nner, though, of course, the description of
this interaction can be improved. Additionally we may consider not
only the ground state of the particle (as happens in the standard
model), but also excited states, in other words, our model does
not necessarily imply a low temperature.

Our final result will be the radius of the void in the form of a
polynomial, in which the coefficients will be related to variables
that could be detected in an expe\-riment, see (\ref{Rad1}) below,
where the radius of the void, $R$, does define indeed a
polynomial, in which the coefficients are functions of parameters
that are under control in an experiment, for instance, $\tau$
(lifetime), $m$ (mass of the particle), etc.
\bigskip
\bigskip

\section{Non--hermitian potentials}
\bigskip
\bigskip

 As usual in the standard model, we treat the ortho--positronium
as a single particle, and refer its wave function with respect to
its center of mass. As mentioned above, we analyze the
relationship between the ortho--positronium lifetime and a
spherical free--volume region with the introduction of a
non--hermitian term in the potential associated with the barrier.
Hence we have that the potential to consider reads

{\setlength\arraycolsep{2pt}\begin{eqnarray} V(r) = 0, ~~r <R,
\label{Pot1}
\end{eqnarray}

\setlength\arraycolsep{2pt}\begin{eqnarray} V(r) = V_0 + i\gamma,
 ~~r\geq R \label{Pot2}.
\end{eqnarray}

Here $\gamma\leq 0$ and $V_0 \geq 0$. Comparing with the standard
model \cite{[7]} we may see that in our case we do not assume a
certain thickness $\Delta R$ associated with the layer defined by
the matrix electrons. At this point it is noteworthy
 to mention that this assumption of the standard model is used also to calculate the wave function
 in such a way that the
 corresponding Schr\"odinger equation has the same form inside the hole and in the region inside
 the polymer defined by
 $R\leq r\leq R + \Delta R$ \cite{[7]}, a questionable fact, since
 this requirement neglects the potential in the aforementioned
 region. Our approach does not need this kind of approximation.

 We may fathom deeper the introduction of the imaginary term in
 the potential noting that in the standard model (see, for
 instance, \cite{[1]}) a homogeneous electron layer with a
 thickness $\Delta R$ must be introduced. Inside this layer takes
 place the annihilation process, in other words, it is a relevant
 parameter of the model. Clearly, $\Delta R$ is an empirical
 variable, and though in the present model we have, at this level,
 also an empi\-rical parameter, namely, $\gamma$,
 there are some important di\-fferences between the two aforementioned approaches that must be underlined.
 Indeed, for instance, in the standard model (see figure 1 of \cite{[1]}) the
 wave function of the positronium (which is, strictly, valid only inside the hole) is extrapolated to the electron layer.
 It is readily seen that in this case the distortion of the wave
 function due to the presence of the layer is not taken into account. In
 addition, since quantization emerges from the introduction of
 boundary conditions, it is also clear that the thickness of the
 layer, in the standard model, does not impinge upon the
 quantization of some of the variables of the positronium, i.e., the energy.

 In our proposal the
 rate of annihilation is considered as an empirical parameter,
 $\gamma$, (this statement can be understood better remembering
 that in the probability conservation equation $\gamma$ plays the
 role of a sink, see (\ref{Con1}) below), and hence we do not need to extrapolate the wave
 function to any region not contained in the hole (therefore we do
 not change the quantization of the energy, for instance). In
 other words, we reduce, compared to the standard model, the
 number of assumptions.

 Additionally, the condition of an infinite potential ba\-rrier is
 not imposed, as is usual in some previous a\-pproaches \cite{[1], [7]}.
 The presence of the matrix of electrons is considered with
 introduction of the extra term, the non--hermitian contribution in the potential
 energy, and it will play the role of a sink for our particles.

Performing the usual procedure we have that the radial part of the
wave function has two different expressions in terms of spherical
Bessel functions

\setlength\arraycolsep{2pt}\begin{eqnarray} y_l(r) = Aj_l(\rho),
~~r<R, \label{Sol1}
\end{eqnarray}

\setlength\arraycolsep{2pt}\begin{eqnarray} y_l(r) =
Bj_l(\tilde{\rho}) + C\eta_l(\tilde{\rho}), ~~r\geq R
\label{Sol2}.
\end{eqnarray}

Here we have that $\rho = \sqrt{{2mE_l\over\hbar^2}}r$, and $\tilde{\rho} =
\sqrt{{2m(E_l - V_0 - i\gamma)\over\hbar^2}}r$.

Quantization of energy appears when we consider the continuity
conditions upon these two expressions, and also upon their logarithmic
derivatives on the surface $r = R$. Clearly, it is possible to consider the case in which the
involved particle is in an excited state $(l\not =0$). In other words,
this approach allows us to consider the possibility of taking into
account, at least partially, the effects of temperature.

If $l = 0$, then the energy reads (approximately)

\setlength\arraycolsep{2pt}\begin{eqnarray} E_0 = V_0 +
\sqrt{{\hbar^4\Omega^2\over 4m^2R^4} - \gamma^2} \label{En1}.
\end{eqnarray}

The parameter $\Omega$, a consequence of the continuity
conditions, is equal to $5.15$. Clearly we may express the
microscopic parameter $\gamma$ as a function of $E_0$, $R$, and
$V_0$.
\bigskip
\bigskip

\section{Ortho--positronium lifetime and free--volume radius}
\bigskip
\bigskip

At this point a link between lifetime and the solutions to the Schr\"odinger
is needed, here we assume that the measured
lifetime comprises two different terms:
\bigskip

1) The time that a particle, with energy $E_0$ and mass $m$, needs
to travel a distance equal to the radius of the free--volume
region.
\bigskip

2) The decay time of an ortho--positronium located inside the polymer.
\bigskip

Concerning these two contributions to the lifetime it has to be
mentioned that in the first part we have a semiclassical approach,
which means that several conditions have to be fulfilled in order
to obtain a good a\-pproximation, for instance, the de Broglie
wavelength of the particle, $\lambda_c = {\hbar\over p}$, has to
be smaller than the radius, i.e., $\lambda_c < R$.

The second part of the present model considers the fact that the
annihilation process takes place inside the polymer at a distance
from the surface $r = R$, which is negligible compared with $R$.

If we consider the continuity equation at any point inside the polymer we
find that

{\setlength\arraycolsep{2pt}\begin{eqnarray}
{\partial\hat{\rho}\over\partial t} + \nabla\cdot {\bf J} =
{2\gamma\over\hbar}\hat{\rho} \label{Con1}.
\end{eqnarray}

Here $\hat{\rho}$ denotes the probability density associated with the
ortho--positronium and ${\bf J}$ the corresponding probability current. At
a fixed point inside the material we have that

{\setlength\arraycolsep{2pt}\begin{eqnarray} \hat{\rho}
=\exp{({2\gamma\over\hbar}t)}f(\vec{r}) \label{Con2}.
\end{eqnarray}

This last expression ($f(\vec{r})$ is a function of the position)
clearly shows that $\gamma$ is related to the decay of the
ortho--positronium when it is not in the void, and in consequence
we may link it with second part of our assumption. In this way we
find that it renders a contribution to the lifetime that goes
(approximately) like $\tau_2 = -{\hbar\over 2\gamma}$.

Hence the lifetime is given by

{\setlength\arraycolsep{2pt}\begin{eqnarray} \tau -{\hbar\over
2\gamma} + \sqrt{{mR^2\over 2E_0}} \label{Ti1}.
\end{eqnarray}

At this point it is noteworthy to comment that we have assumed,
from square one, that the positronium is always localized inside
the void. In other words, we do not consider a modification of a
situation in which the positronium may escape from the hole, a
case already studied \cite{[10]}.
\section{Theoretical predictions}
\bigskip
\bigskip

It is readily seen that in our last expression, if we seek to
obtain information concerning the size of the void, then we must
know the terms which are available in an experiment. In some cases
\cite{[9]} the annihilation rate inside the electron layer is an
experimental result, in our case this means that we may know the
value of $\gamma$. Bearing this in mind we may cast (\ref{En1})
and (\ref{Ti1}) as a polynomial equation for the radius

\begin{eqnarray}
{m^4\over (\tau +{\hbar\over 2\gamma})^4}R^8  -4{m^3V_0\over (\tau
+{\hbar\over 2\gamma})^2}R^6 \nonumber\\
+ 4m^2(V_0^2 + \gamma^2)R^4 = \hbar^4\Omega^2. \label{Rad1}
\end{eqnarray}

Let us now introduce the following definition

\begin{eqnarray}
R = \sqrt{{4\tau^2V_0\over m}}r. \label{Rad2}
\end{eqnarray}

Then we may rephrase (\ref{Rad1}) in terms of the dimensionless
parameter $r$

\begin{eqnarray}
r^8 -r^6 + {1\over 2}r^4 = {\hbar^4\Omega^2\over 4^4\tau^4V_0^4}.
\label{Rad13}
\end{eqnarray}

 This last expression embodies the main
result in the present work. In order to resort to (\ref{Rad1}) as
a model linking the lifetime and void size we require, in
addition, the value of $\gamma$ and $V_0$. As previously
mentioned, in some materials the annihilation rate inside the
electron layer is known. Forsooth, in the case of positronium the
annihilation rate inside the electron layer is the spin--averaged
rate for the ortho--Ps and para--Ps, which reads $2ns^{-1}$
\cite{[9]}. In our case, this implies that

\begin{eqnarray}
{1\over 2}ns = - {2\gamma\over\hbar}. \label{Ann1}
\end{eqnarray}

It is readily seen that a theoretical prediction requires a deeper
knowledge of the electron layer, to wit, the value of $V_0$. It is
also clear that the evaluation of this parameter lies outside the
realm of the present approach. Nevertheless, in order to have an
idea of the possibilities of the present model let us now assume
that $V_0$ has an order of magnitude similar to the contribution
of one atom to the internal energy of a solid (within the Einstein
model \cite{[11]}) in the limit in which the temperature of the
solid is much higher than the characteristic Einstein temperature.
Therefore

\setlength\arraycolsep{2pt}\begin{eqnarray} V_0 =\kappa T.
\label{Ein1}
\end{eqnarray}

The introduction now of $T \sim 300 K$, in the case of the
lifetime of an $\alpha$--cage, of the so--called MS--4A zeolite
\cite{[8]}, the one reads $\tau \sim 5.9$ns, renders

\begin{eqnarray}
R = 0.42nm. \label{Radt}
\end{eqnarray}

The experimental value for this case reads \cite{[8]}

\begin{eqnarray}
R = 0.57 nm. \label{RadE}
\end{eqnarray}

In other words, our model provides a radius that has the same
order of magnitude than the experimental measurement.
\section{Conclusions}
\bigskip
\bigskip

The case of an ortho--positronium immersed in a sphe\-rically
symmetric potential wall, in which the potential barrier contains
a non--hermitian term, has been consi\-dered as a model to obtain
a correlation between lifetime and free--volume regions in some
polymers.

This approach allows us to take into account the effects of
temperature, i.e., we may consider in expressions (\ref{Sol1}) and
(\ref{Sol2}) the case $l\not = 0$, a fact that implies that the
particle is not in its ground state. Indeed, let us now suppose
that $l = 1$, then

\setlength\arraycolsep{2pt}\begin{eqnarray} y_1(r) =
A\Bigl[{\sin(kr)\over (kr)^2} - {\cos(kr)\over (kr)}\Bigr], ~~r< R
\label{Sol3}
\end{eqnarray}

\setlength\arraycolsep{2pt}\begin{eqnarray} y_1(r) =
B\Bigl[{\sin(\tilde{k}r)\over (\tilde{k}r)^2} -
{\cos(\tilde{k}r)\over (\tilde{k}r)}\Bigr] \nonumber\\
+ C\Bigl[{\cos(\tilde{k}r)\over (\tilde{k}r)^2} +
{\sin(\tilde{k}r)\over (\tilde{k}r)}\Bigr], ~~r\geq R.
\label{Sol4}
\end{eqnarray}

Proceeding in the same way as in the first case we find that

\setlength\arraycolsep{2pt}\begin{eqnarray} E_1 = V_0 +
\sqrt{{\hbar^4\hat{\Omega}^2\over 4m^2R^4} - \gamma^2}.
\label{En2}
\end{eqnarray}

The parameter $\hat{\Omega}$, a consequence of the continuity conditions, is equal to $17.49$.
In a first approximation
we have that the present models predicts

\setlength\arraycolsep{2pt}\begin{eqnarray} E_1 - E_0=
13{\hbar^2\pi^2\over mR^2}. \label{En3}
\end{eqnarray}

In the standard model this energy difference \cite{[8]} reads
$\Delta E= 3{\hbar^2\pi^2\over mR^2}$, in other words, our
prediction is 13/3 times larger than the one stemming from the
standard model.

Clearly, one of the mayor objections to the standard model,
spherical symmetry, is here also present, and, as it has already
been pointed out \cite{[7]}, even ellipsoidal sy\-mmetry is
unlikely to represent the real situation. Ne\-vertheless, the
simple case that we have considered here allows us to obtain
analytical results (we do not have to resort to molecular dynamics
simulations in order to improve the standard model \cite{[7]}).

Additionally we avoid a second drawback of the standard model,
closely related to the presence of a layer of homogeneous density,
and thickness $\Delta R$, defined by the matrix electrons. Indeed,
the introduction of this layer entails also an a\-dditional
condition, to wit, the wave function inside the layer is taken
from the solution that emerges inside the spherical hole
\cite{[1], [7]}, a situation that can not be correct. In our case,
we do not need introduce this assumption.

The effects of the polymer are taken into account with the
introduction of $\gamma$, a microscopic parameter that can be
straightforwardly connected with an experimental output
\cite{[9]}, and with $V_0$. Resorting to a very rough
a\-ssumption, expression (\ref{Ein1}), has rendered an order of
magnitude for the radius of the void that matches with the
experimental readout. Of course, a more precise prediction demands
a more physical value for $V_0$ (than the one here employed),
nevertheless (\ref{Radt}) and (\ref{RadE}) show that the present
model could be a useful one.

To conclude, it is already a known result \cite{[12]} that the
holes show a significant asphericity. Hence, a further case in the
context of the present work is related to the po\-ssibility of
improving the standard model trying to gene\-ralize the present
approach to situations lacking spherical symmetry, for instance,
considering first ellipsoids, as in \cite{[1]}, and introducing
non--hermitian terms in the potential energy. Though, as
\cite{[1]} clearly shows, we must resort to numerical methods.
\bigskip
\bigskip

\Large{\bf Acknowledgments}\normalsize
\bigskip

This research was partially supported by CONACYT Grant 42191--F.
The author would like to thank A.A. Cuevas--Sosa and A.
Mac\'{\i}as for useful discussions and literature hints.

\end{document}